\documentstyle[12pt,a4,epsf]{article}

\newcommand\order{\mathop{\cal O}\nolimits}

\newcommand{\lsim}{\stackrel{\lower.7ex\hbox{$<$}}{\lower.7ex\hbox{$\sim$}}}
\def\leftrightaccent#1{\shortstack[b]{$\scriptstyle\leftrightarrow$\\$#1$}}
\newcommand\GeV{\mbox{ GeV}}

\newcommand\PB{\mbox{pb}}
\newcommand\inp{{\cdot}}
\newcommand\kmax{{k_{\rm max}}}
\def\setepsfscale#1{\def\epsfsize##1##2{#1##1}}

\begin{document}


\thispagestyle{empty}
\setcounter{page}{0}

\begin{flushright}
PSI-PR-94-16\\
May 1994
\end{flushright}

\vspace{\fill}

\begin{center}
{\Large\bf EEWW: a generator for $e^+e^- \to W^+W^-$ including one-loop and
leading photonic two-loop corrections}\\[1.5cm]
\begin{tabular}[t]{c}
\large Jochem~Fleischer$^a\strut$\\
\large Fred~Jegerlehner$^b\strut$\\
\large Karol Ko\l odziej$^c\strut$\\
\large Geert~Jan van~Oldenborgh$^b\strut$\\[1cm]
{\it$^a$ Fakult\"at f\"ur Physik, Universit\"at Bielefeld,
Universit\"atstra\ss e 25,}\\{\it D-33615 Bielefeld, Germany}\\
{\it$^b$ Paul Scherrer Institut, CH-5232 Villigen PSI, Switzerland}\\
{\it$^c$ Sektion Physik der Universit\"at M\"unchen, Theresienstra\ss e 37,}\\
{\it D-80333 M\"unchen, Germany}\\
\end{tabular}
\end{center}

\vspace{\fill}

\begin{abstract}\noindent
We describe a generator for the process $e^+e^-\to W^+W^-$ including all
one-loop and leading log photonic two-loop contributions.  It includes
polarization of the beam and $W$ bosons, and the possibility to study the
effect of anomalous couplings.
\end{abstract}

\vspace*{\fill}
\newpage


\section{Introduction}

The computation of the one loop $W$ pair production cross section in
$e^+e^-$ collisions has been performed a long time ago
\cite{Lemoine&Veltman,BoehmWW,Fleischer&JegerlehnerWW}.  The effects of hard
radiation have been added to the last two calculations
\cite{Wim&Karol&ThomasW,Jochem&Fred&KarolW}.  These results agree at the order
of $10^{-3}$ or better \cite{Jochem&Fred&KarolW}, except at high energies in
the forward direction, which is probably attributable to numerical
instabilities.  With the advent of LEP II there is a need for an event
generator for $W$ pair production at this level.\footnote{After writing this
article it has come to our attention that a similar effort has been reported
before \cite{MartinezWMWC}.}  We present here such a
generator based on the computations of Ref.\ \cite{PolarizedWW}.  It includes
the full one-loop matrix element, leading logarithm two-loop initial state
bremsstrahlung, polarization effects and anomalous couplings.  The main
limitation is that we assume the $W$ bosons to be stable particles.  Work to
remove this limitation is in progress. We nevertheless think that the present
version is useful in several respects: it includes the interesting part of the
full process $e^+ e^- \rightarrow W^+ W^- \rightarrow$ four fermions and thus
allows to study what we may learn from $W$--pair production under idealized
circumstances. The missing part only involves well known physics, namely, the
decays via the well established charged current interaction. The decays thus
just serve as polarization analyzers for the $W$'s. Since a full $O(\alpha)$
calculation of the observed process $e^+ e^- \rightarrow$ four fermions is a
major enterprise, we think it will be useful for tests of parts of more
complete calculations.

We first recapitulate the construction of an event generator, then discuss the
polarization, initial state collinear bremsstrahlung and anomalous couplings.
In the appendices technical details about the installation and sample input
and output are given.


\section{Event generator}

The event generator presented here is based on the calculations
\cite{Fleischer&JegerlehnerWW,Jochem&Fred&KarolW,PolarizedWW}.  These,
however, have all been presented as total and differential cross sections, or
cross sections with certain cuts in the photon energy and the angle between
the photon momentum and the beam axis.  In the present work we
present an event generator, i.e., a program which will give configurations
$W^+W^-(\gamma)$ with probability proportional to the contribution to the
total cross section.  This allows experimental studies to be done including
arbitrary cuts and detector capabilities.  In this section we describe how
the conversion was performed.

The last step to an event generator is simple: if the cross section is written
as an integral over a positive function $f(\vec{x})$, with $\vec{x}$ a point
in a hypercube $0<x_k<1$, and $f_{\rm max}$ the maximum value of the function
in this hypercube, event generators are given by the representation of the
cross section
\begin{equation}
\label{eq:sum}
    \sigma = \int\!\!d\vec{x} \, f(\vec{x})
        \approx \frac{1}{N} \sum_{i=1}^N f(\vec{x}_i)
        \approx \frac{f_{\rm max}}{N} \sum_{i=1}^N
            \theta\Bigl(\frac{f(\vec{x}_i)}{f_{\rm max}} - r_i\Bigr)
\;,
\end{equation}
where the $\vec{x}_i$ are random points in the hypercube and the $r_i$ random
numbers between 0 and 1; $N$ is the number of function evaluations used.  The
first sum corresponds to a {\em weighted\/} event generator, the second one to
a {\em weight-one\/} event generator --- all events correspond to the same
probability.  Both options are incorporated in {\tt EEWW}, the first using the
adaptive Monte Carlo routines {\tt vegas} \cite{LepageVEGAS}, the second one
using non-adaptive algorithms {\tt simplemc} and {\tt axmc}.  Experimental
cuts and efficiencies can now easily be incorporated in the sums as
$\theta$-functions, possibly involving more random numbers.  For parton-level
calculations a weighted Monte Carlo is generally more efficient, as the main
cost is the evaluation of the matrix element.  When a detector simulation is
included, however, this part will take much more time, so a weight-one
generator is generally used.  The efficiency of the weight-one generator is
given by
\begin{equation}
    \epsilon = \int\!\! d\vec{x} \, \frac{f(\vec{x})}{f_{\rm max}}
        \approx \frac{1}{N} \sum_{i=1}^N
            \theta\Bigl(\frac{f(\vec{x}_i)}{f_{\rm max}} - r_i\Bigr)
\;,
\end{equation}
i.e., the average number of events generated per function evaluation.

The maximum value of $f(\vec{x})$ may not be known beforehand.  We obtain an
estimate by sampling the function at $10^{1+n/6}$ points for an
$n$-dimensional function, and taking 20\% more than the largest value found.
This is adequate for a reasonably flat function.  If during the generation a
function value is found which is larger than the assumed maximum value, Eq.\
(\ref{eq:sum}) is adjusted as
\begin{equation}
    \theta\Bigl(\frac{f(\vec{x}_i)}{f_{\rm max}} - r_i\Bigr)
    \to
    \mbox{int}\frac{f(\vec{x}_i)}{f_{\rm max}} +
        \theta\Bigl(\frac{f(\vec{x}_i)}{f_{\rm max}} -
        \mbox{int}\frac{f(\vec{x}_i)}{f_{\rm max}} - r_i\Bigr)
\;,
\end{equation}
where $\mbox{int}(x)$ denotes the largest integer smaller than $x$.  This
means that the event is accepted as many times as is necessary to obtain the
correct integral.  As a result, not all events are independent, and
statistical fluctuations may be larger than expected.  Note that
$f_{\rm max}$ is not readjusted to reflect the new maximum found.

For our purpose we need to write the one loop cross section as an integral
over a positive function on a hypercube.  The hard radiation is already in
this form:
\begin{eqnarray}
\label{eq:hard}
    \sigma_1^{\rm H} & = & \frac{1}{2s}
        \int_0^{2\pi}\!\!d\phi_W
        \int_{-1}^{+1}\!\!d\cos\theta_W
        \int_{k_0 E_e}^{\kmax E_e}\!\!dE_\gamma
        \int_0^{2\pi}\!\!d\phi_\gamma
        \int_{-1}^{+1}\!\!d\cos\theta_\gamma
        J |{\cal M}_1^{\rm H}|^2
\;,
\end{eqnarray}
with $J$ the Jacobian defined in Ref.\ \cite{Jochem&Fred&KarolW} and $E_e$ the
beam energy $E_e=\sqrt{s}/2$.  $k_0$ is the minimum fraction of the beam
energy the photon is allowed to have.  The mappings from the $x_i$ to the
variables in Eq.\ (\ref{eq:hard}) are also described in this reference; these
map away the infrared $1/E_\gamma$, the collinear $1/(E_e \pm
p_e\cos\theta_\gamma)$, and the $t$-channel $1/t$ peaks of the matrix element
squared.

The Born, virtual and soft contributions to the total cross section have the
same form without photon integrals, so the integral can be taken the same with
a flat behaviour in these dimensions.  These two integrals are combined by
introducing an extra (sixth) integration variable
\begin{equation}
\label{eq:hardother}
    \sigma = \sigma_0 + \sigma_1^{\rm V+S} + \sigma_1^{\rm H}
        = \int_0^1\!\!dx_6 \, \Bigl\{\frac{\theta(a-x_6)}{a}
            \left(\sigma_0 + \sigma_1^{\rm V+S}\right) +
            \frac{\theta(x_6-a)}{1-a}\sigma_1^{\rm H} \Bigr\}
\;,
\end{equation}
with $0<a<1$ a tunable parameter.
The distinction between hard and soft radiation is arbitrary; the only demand
is that the cutoff $k_0$ is much smaller than unity to validate the eikonal
approximation used in the soft bremsstrahlung integrals.  We use this freedom
to make the first term as small as possible without making it negative; a
suitable choice is
\begin{equation}
\label{eq:k0}
    k_0 = \frac{\sqrt{s}-A}{B}\exp\frac{\displaystyle -{\pi \over
        2\alpha} - {\pi^2\over6} + 1 - {3\over4}\log(m_e^2/s)}{\log(m_e^2/s)-1}
        \;.
\end{equation}
The parameters $A = 130\GeV$, $B = 35\GeV$ in the $\alpha$ scheme ($B=9\GeV$
in the $G_\mu$ scheme) and $a = \sqrt{s}/1000\GeV$ below $\sqrt{s}=350\GeV$,
$a=0.35$ above have been found by trial and error.  At LEP II energies this
gives $(\sigma_0 + \sigma_1^{\rm V+S}) / \sigma_1^{\rm H} \sim 1/7$.  As  the
virtual matrix element takes much longer to evaluate than the bremsstrahlung
amplitude it is advantageous to keep this ratio as small as possible.  Using
these values we obtain a reasonably flat function $f(\vec{x})$.  The
efficiency, without any cuts, is about 40\%.  This translates to about 5
events per second on a workstation.

Note that the resulting soft photon spectrum, which seems to continue down to
below 1 MeV, can not be trusted below a few times the $W$ width because of the
narrow width approximation.


\section{Polarization}

Building on the calculations presented in Ref.\ \cite{PolarizedWW} the event
generator has many possibilities for studying the effect of different
polarizations of the initial and final state.

The state of the beam is characterized by the electron and positron density
matrices in the helicity frame
\begin{eqnarray}
    \rho_- & = & \frac{1+\vec{\sigma}\inp\vec{P}}{2} =
        \left(\begin{array}{cc} 1+P_L & P_Te^{-i\phi} \\
                                P_Te^{i\phi} & 1-P_L \end{array}\right) \\
    \rho_+ & = & \frac{1+\vec{\sigma}\inp\vec{P}'}{2} =
        \left(\begin{array}{cc} 1+P_L' & P_T'e^{i\phi'} \\
                                P_T'e^{-i\phi'} & 1-P_L' \end{array}\right)
\end{eqnarray}
where $\vec{P} = (P_T\cos\phi, P_T\sin\phi,P_L)$, $P_L$ is the longitudinal
polarization and $P_T$ the magnitude of the transverse polarization of the
electron; $P_L'$, $P_T'$ of the positron.  The polarization vector of the
electron (positron) is pointed in a direction $\phi$ ($\phi'$) upwards
(downwards) from the outward direction.  For the natural polarization in a
storage ring $\phi = \phi' = 90^\circ$.  Special cases are no polarization
($P_L=P_T=0$) and longitudinal polarization ($P_T=P_T'=0$, $P_L=\pm1$,
$P_L'=0$ for right- and left-handed polarization); these can be chosen
separately.  In the limit that the electron is taken massless the amplitudes
for like-handed electrons and positrons ($P_L=P_L'$) are zero.  The matrix
element is then given by
\begin{eqnarray}
    \overline{|M|^2} & = &{1\over4} \Bigl[ (1-P_L P'_L)
        \bigl(|M_+|^2+|M_-|^2 \bigr) + (P_L-P'_L)\bigl(|M_+|^2-|M_-|^2\bigr)
\nonumber\\&&\hspace*{-1cm}\mbox{} + (2 P_T P'_T) \left(
        \cos (\phi-\phi'-2\phi_W)\;{\rm Re} (M_+ M^*_-) +
        \sin (\phi-\phi'-2\phi_W)\;{\rm Im} (M_+ M^*_-) \right)
                \Bigr]
\;.
\end{eqnarray}
A detailed discussion of the effect of the choices is given in Ref.\
\cite{PolarizedWW}.\footnote{Note that in table V of this reference
$P_T=P_T'=1$ is taken, like in the definition (21) but unlike the suggestion
in the text, and a factor two has been included in the definition of the
asymmetry.  The typo corrections $\cos\phi\to\phi$ in (21) and
$\Delta\phi=\phi-\phi'-2\phi_W$ after (17) are obvious.}

The $W$ bosons can have three polarization states: 2 transverse and 1
longitudinal.  One can either choose to average over these in the matrix
element (`unpolarized' $W$'s), or generate them separately (`polarized').  In
the latter case the polarization vector is made available, so that the
subsequent decay of the $W$ can take this information into account.   The
other two possibilities are to generate only longitudinal or transverse
$W^\pm$ bosons.

The generation of different polarization states is implemented by converting
the sum over final states into an integral as
\begin{equation}
\label{eq:sumpol}
    \sum_{\lambda^\pm=-1,0,+1} \sigma(\lambda^+,\lambda^-) =
        \int_0^1 \!\! dx \sum_{i=0}^8
        \frac{\theta(x-c_i)\theta(c_{i+1}-x)}{c_{i+1}-c_i}
        \sigma(\lambda^+_i,\lambda^-_i)
\;,
\end{equation}
with $\lambda^+_i = \mbox{int}(i/3)-1$, $\lambda^-_i = \bmod(i,3)-1$.  The
cutoff values $c_i$ are chosen according to the lowest order matrix element
squared at the same $W^-$ angle:
\begin{equation}
    c_{i+1} - c_i = \frac{|M_0(\lambda^+_i,\lambda^-_i)|^2}
            {\sum_{j=0}^8|M_0(\lambda^+_j,\lambda^-_j)|^2}
\;.
\end{equation}
The integration variable to make the discrete choice (\ref{eq:sumpol}) is
taken to be the rescaled version of the one used for the choice between the
hard and other corrections (\ref{eq:hardother}).


\section{Initial state collinear bremsstrahlung}

The largest corrections in the one-loop calculation are caused by initial
state collinear bremsstrahlung.  It is therefore natural to seek a way to
include these to higher order.  The leading log terms are well known
\cite{LeidenEemumu2loop}.  They were included in
Ref.~\cite{Jochem&Fred&KarolW} by replacing the lowest order term
$\sigma_0(s)$ by
\begin{equation}
\label{eq:struct}
    \sigma^{\rm ini}(s) = \int_0^\kmax dk \,\rho_{\rm ini}(k) \,
        \sigma_0\bigl(s(1-k)\bigr)
\;,
\end{equation}
with the function $\rho_{\rm ini}$ given by
\begin{equation}
\label{eq:rho}
    \rho_{\rm ini}(k) = \beta k^{\beta-1}\Bigl(1 + \delta_1^{\rm V+S} +
        \delta_2^{\rm V+S} \Bigl) + \delta_1^{\rm H} + \delta_2^{\rm H}
\;,
\end{equation}
with $\beta = 2\frac{\alpha}{\pi}\bigl(L-1\bigr)$ and $L = \log(s/m_e^2)$ the
large collinear logarithm.  The terms $\delta$ denote the infrared finite
parts of the leading one- and two-loop corrections.  Explicit expressions are
given in Refs \cite{LeidenEemumu2loop,Jochem&Fred&KarolW}, except that we use
$\delta_1^{\rm H} = {\alpha\over\pi}\bigl((L-1)(k-2) + k\bigr)$, which differs
by the last (non-leading) term.\footnote{Numerically, the difference is less
than 0.1\%.}  However, one has now included the one-loop leading corrections
twice; this is corrected for by subtracting the unexponentiated
$\order(\alpha)$ contribution from this formula:
\begin{equation}
    \sigma^{\rm double} = \left(\beta\log{k_0} + \delta_1^{\rm V+S}\right)
        \sigma_0(s) + \int_{k_0}^\kmax dk \Bigl( {\beta \over k} +
        \delta_1^{\rm H} \Bigr) \sigma_0\bigl(s(1-k)\bigr)
\;,
\end{equation}
with $k_0$ an arbitrary (small) cutoff.

This procedure, however, is not suited for an event generator.  First, a
separate structure function has to be assigned to each incoming particle
\cite{KuraevFadinStruct,AltarelliMartinelliStruct,TrentadueStruct,KleissIni};
these are identical to $\rho_{\rm ini}$ with $\beta\to\beta/2$;
for simplicity we will keep the single integral notation.  A more subtle point
is that the structure function (\ref{eq:rho}) has been computed by solving QED
evolution equations after integrating out the angles of the extra photon over
all phase space.  This inclusive approach is not appropriate for an event
generator.  We solve this by dividing phase space into two regions: inside a
cone of opening angle $\theta_c$ around the beam direction one declares all
photons collinear (and makes an inclusive measurement), outside it one
measures the angles (and measures exclusively).  Only in the phase space
inside this cone ($\theta_c$ is assumed to be much larger than $m_e^2/s$) we
use the exponentiated two-loop structure function.  Outside this cone the
$\order(\alpha)$ result is used, with the possibility of extra collinear
photons.  The two limiting cases are $\theta_c = 0$, which gives a strict
$\order(\alpha)$ behaviour, and $\theta_c = \pi$, which reproduces the totally
integrated cross section computed before.\footnote{The region $\pi/2 \le
\theta_c < \pi$ is redundant from an experimentalists' point of view, but the
collinear terms have to be integrated over all angles to obtain the full
result.}  Integrating only over the angles inside the cone amounts to
replacing $s$ by a scale $\mu^2 = s(1-\cos\theta_c)/2$ in the collinear large
logarithms: $L = \log(\mu^2/m_e^2)$.  A calculation of the $p_T$ dependence of
the structure functions has not been done yet and is beyond the scope of this
article.

A technical problem is that adding and subtracting parts does not give the
positive definite integrand needed for an event generator.  To obtain this we
rewrite the total cross section as
\begin{eqnarray}
    \sigma & = & \sigma^{\rm ini} + \sigma_1^{\rm V+S} + \sigma_1^{\rm H} -
        \sigma^{\rm double}
\nonumber\\
    & = & \int_0^\kmax dk \,\rho_{\rm ini}(k) \, \sigma_0\bigl(s(1-k)\bigr)
        + \Bigl[\sigma_1^{\rm V+S}(s) - (\beta\log{k_0} + \delta_1^{\rm V+S})
            \sigma_0(s) \Bigr]
\nonumber\\&&\quad
        + \Bigl[\sigma_1^{\rm H}(s) - \int_{k_0}^\kmax dk \Bigl(
            {\beta \over k} + \delta_1^{\rm H} \Bigr)
            \sigma_0\bigl(s(1-k)\bigr) \Bigr]
\;.
\label{eq:subtractLL}
\end{eqnarray}
The terms in square brackets are of order $\alpha$ and do not contain any
large collinear logarithms; we can therefore add them to the first integral
without introducing terms of order $\alpha^2 L$:
\begin{equation}
    \sigma = \int_0^\kmax dk \,\rho_{\rm ini}(k) \, \left(
        \sigma_0 + \tilde{\sigma}_1^{\rm V+S} + \tilde{\sigma}_1^{\rm H}
        \right) \bigl(s(1-k)\bigr)
\;,
\end{equation}
with $\tilde{\sigma}_1^i$ the differences in square brackets in
Eq.~(\ref{eq:subtractLL}).  The virtual and soft one is easily computed, to
subtract the leading log from the hard radiation we have to reintroduce the
$\cos\theta$ integral as (see, e.g., \cite{epHiggs})
\begin{equation}
    \int_{k_0}^\kmax dk \, {\beta \over k} = {\alpha \over \pi}
            \int_{k_0E_e} {dE_\gamma \over E_\gamma} \Biggl(
        \int_{-1}^{-\cos\theta_c} \! d\cos\theta
            {p_e \over E_e + p_e\cos\theta}
        + \int_{\cos\theta_c}^{1} \! d\cos\theta
            {p_e \over E_e - p_e\cos\theta} \Biggr)
\label{eq:cost}
\end{equation}
This can be subtracted from the hard radiative integral over a 3-particle
phase space if one adds a $\phi$ integral (which is trivial for $P_T=0$,
otherwise the same $\phi$-dependence is taken as the lowest order).  A
similar approach is followed for the double pole terms; here the upper
integration boundary is of order $m_e^2/s(1-\cos\theta_c)$ which can be
neglected when $\theta_c\ll m_e^2/s$.  The final result is just the collinear
limit used already in Refs~\cite{KleissMass,Jochem&Fred&KarolW}, but now
applied for all angles up to $\theta_c$ in both the forward and backward
direction.

Unfortunately there is no reason for the integrand of $\tilde{\sigma}_1^{\rm
H}$ to be positive definite; indeed, in the collinear limit it is zero.
Recalling that within the collinear cone we do not make a distinction between
different collinear photons we add a fraction of the Born cross section to
restore positivity.  We take
\begin{eqnarray}
    \tilde{\tilde{\sigma}}_1^{\rm H} & = & \sigma_1^{\rm H} -
        \frac{\alpha}{\pi} \int_{k_0}^\kmax \!\!dk
        \Bigl(\int_{-1}^{-\cos\theta_c} \!\! + \int_{\cos\theta_c}^{1}\Bigr)
        d\cos\theta \Bigl( \frac{1 + (1-k)^2}{k} \frac{1}{p_\pm \inp p_\gamma}
        + (1-k)\frac{m_e^2}{(p_\pm \inp p_\gamma)^2} \Bigr)
\nonumber\\&&\quad\times
        \Bigl[ \frac{\frac{1}{1-k}d\sigma_0\bigl((1-k)s\bigr)}{d\cos\theta} -
        \frac{d\sigma_0(s)}{d\cos\theta} \Bigr]
\\
    \tilde{\tilde{\sigma}}_1^{\rm V+S} & = & \sigma_1^{\rm V+S} -
        \frac{\alpha}{\pi} \left( 2(L-1)\log\kmax + \bigl(\frac{3}{2}
        - 2\kmax\bigr)L + \frac{\pi^2}{3} - 2 + 2\kmax + \frac{1}{2}\kmax^2
        \right) \sigma_0(s)
\nonumber\\
\;,
\end{eqnarray}
where $p_\pm$ is the positron (electron) momentum in the first (second)
$\cos\theta$-integral.  Note that $\kmax$ depends on the angle between the
$W^-$ and the incoming electron.  As the expression for the hard
bremsstrahlung is dominated by the added Born term we generate an event with
only collinear photons in this region.


\section{Anomalous couplings of the $W$ boson}

One of the main purposes of LEP II being the study of the $WW\gamma$ and $WWZ$
interactions we allow for the inclusion of non-standard couplings at these
vertices \cite{BlondelinZPhysics,GaemersGounaris}.  We parametrize these by
the CP-invariant effective interaction Lagrangean for the $WWV$ interaction
($V = \gamma$ or $Z$)
\begin{eqnarray}
    {\cal L}_{\rm SM}^V & = & iC_V\left\{ V^\nu\bigl(W^+_{\mu\nu}W^{-\mu} -
        W^-_{\mu\nu}W^{+\mu}\bigr) + V^{\mu\nu}W^+_\mu W^-_\nu \right\}
\\
    \Delta{\cal L}_{\rm eff}^V & = &
        iC_V\delta_V\left\{ V^\nu\bigl(W^+_{\mu\nu}W^{-\mu} -
            W^-_{\mu\nu}W^{+\mu}\bigr) + V^{\mu\nu}W^+_\mu W^-_\nu \right\}
        + iC_V\delta\kappa_V V^{\mu\nu}W^+_\mu W^-_\nu
\nonumber\\&&\mbox{}
        + iC_V\frac{\lambda_V}{m_W^2}V^\mu{}_\nu W^{+\nu\rho}W^-_{\rho\mu}
        + C_V\xi_V\Bigl(\partial^\mu\tilde{V}^{\nu\rho}\Bigr)
            \left\{W_\mu^+ \leftrightaccent{\partial}_\nu W^-_\rho
                 + W_\mu^- \leftrightaccent{\partial}_\nu W^+_\rho \right\}
\;.
\label{eq:defano}
\end{eqnarray}
Electromagnetic gauge invariance forces $\delta_\gamma = 0$.  This effective
Lagrangean is related to the other widely used form \cite{ZeppenfeldHagiwara}
by $g_1^V = 1 + \delta_V$, $\kappa_V = 1 + \delta_V + \delta\kappa_V$ and
$f_5^V = \xi_V s/m_W^2$.  Whereas CP-violating terms are known to be small, we
see no reason to neglect the parity-violating form factors $\xi_V$.

The inclusion of these anomalous couplings in a one-loop expression causes
problems, as the resulting theory is non-renormalizable.  We assume that the
anomalous couplings are small perturbations of the standard model, and that
the leading effects enter only in the Born term.  They are therefore not
included in the $\order(\alpha)$ corrections, in particular not in the hard
(non-collinear) bremsstrahlung.
The usage of a different expression for the Born term in the soft
bremsstrahlung will upset the validity of Eq.\ (\ref{eq:k0}) which assures
the positivity of the integrand.  This only causes negative events when the
anomalous couplings decrease the Born matrix element squared, which, due to
the violation of the gauge cancellations, happens only in the backward
direction ($\cos\theta_W \approx -1$).  We therefore increase the cutoff by a
factor $1/x_5^2$, where $x_5$ is the random variable which is mapped to
$\cos\theta_W$.  In case of problems one can rescale $k_0$ by an arbitrary
factor.  At high energies, for large anomalous couplings and with extra
initial state collinear radiation it may not be possible to avoid negative
points.


\section{Anomalous couplings and loop effects}

We have compared the effects of anomalous couplings of the $WWZ$ vertex to the
one-loop corrections in the total cross section and the angle of the $W^-$
boson.  One should note that the effects of the one-loop corrections is to
entirely different amplitudes than those affected by anomalous couplings.  The
effective one-loop contributions to the anomalous couplings defined in Eq.\
(\ref{eq:defano}) are of order $\alpha/\pi$, i.e., a few times $10^{-3}$.
This is completely negligible at the precision expected at LEP II.  The large
one-loop contributions therefore originate in different form factors, like the
ones associated with the $t$-channel graphs.  These also affect simple
distributions like $d\sigma/d\cos\theta_W$.

\begin{figure}[htb]
\centerline{
\unitlength .5bp
\setepsfscale{.5}
\begin{picture}(490,524)(25,0)
\put(0,0){\strut\epsffile{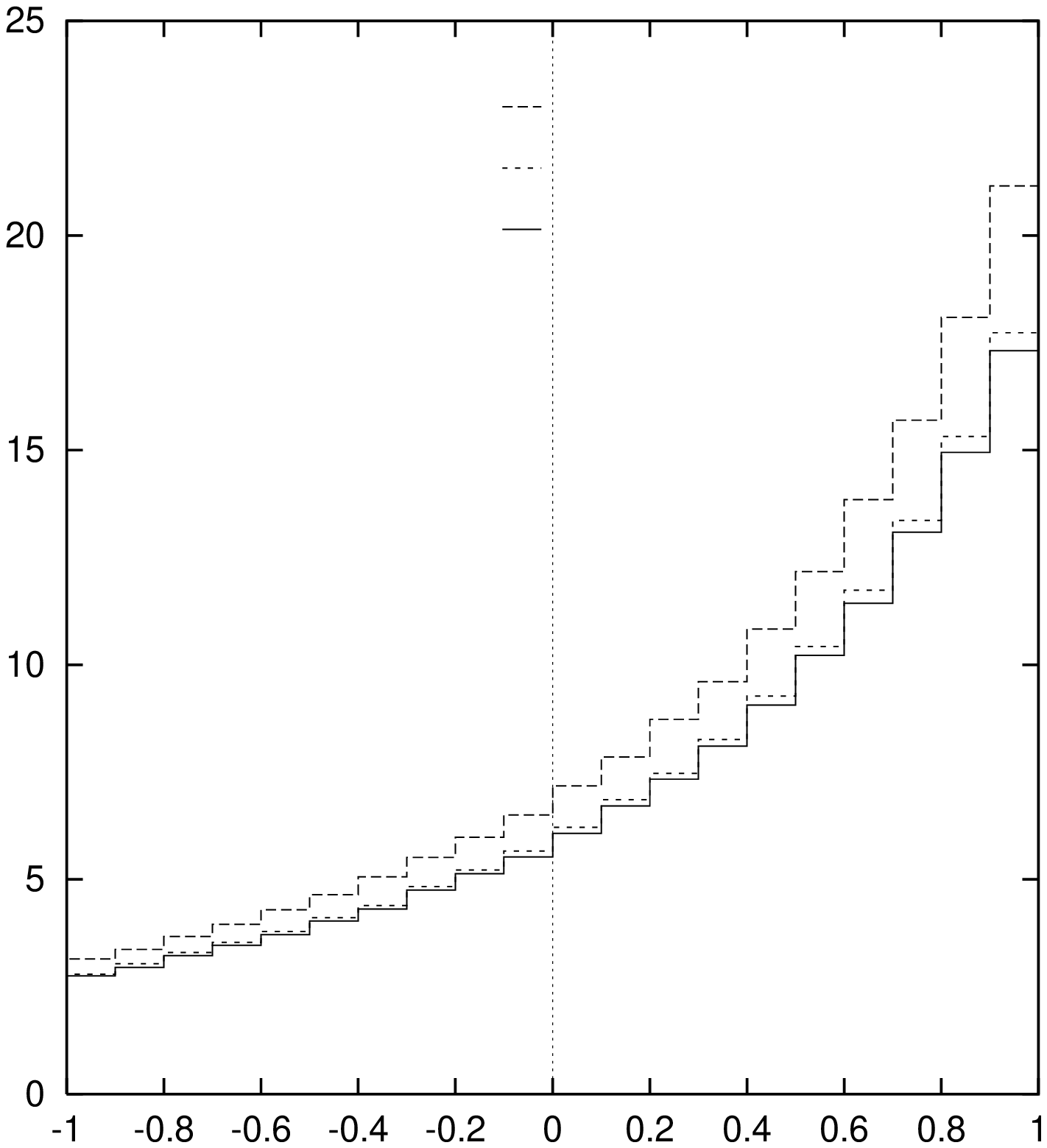}}
\put(480.9,0){\makebox(0,0)[rb]{$\cos\theta_W$}}
\put(80,498.9){\makebox(0,0)[bl]{$\displaystyle \frac{d\sigma}{d\cos\theta_W}
 \;[\PB]$}}
\put(262.8,431.3){\makebox(0,0)[rb]{\footnotesize strict Born}}
\put(262.8,407.3){\makebox(0,0)[rb]{\footnotesize structure function}}
\put(262.8,383.3){\makebox(0,0)[rb]{\footnotesize full one loop}}
\end{picture}
\begin{picture}(504,524)(25,0)
\put(0,0){\strut\epsffile{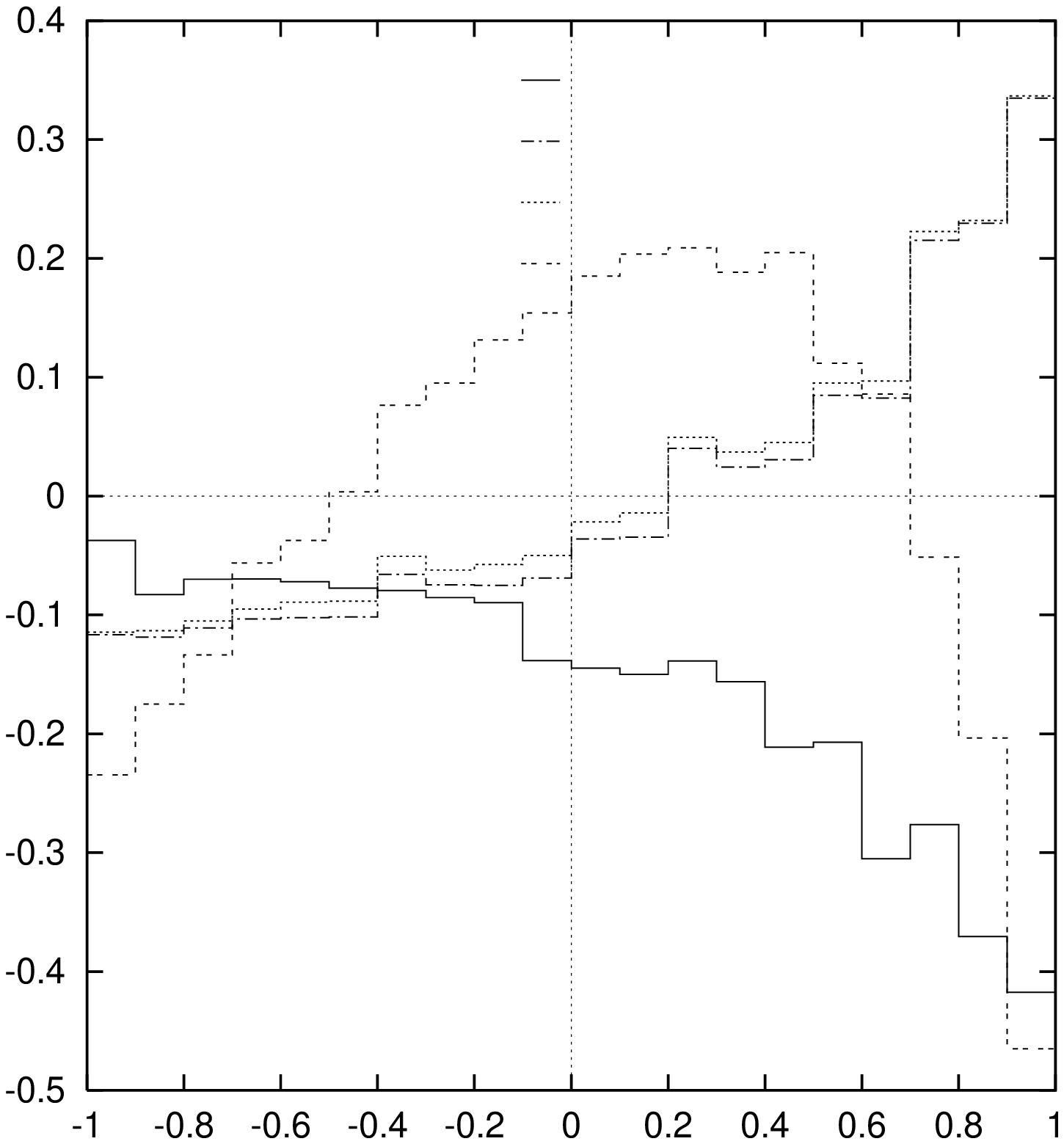}}
\put(480.9,0){\makebox(0,0)[rb]{$\cos\theta_W$}}
\put(80,498.9){\makebox(0,0)[bl]{$\displaystyle
	\frac{d(\sigma - \sigma_{\rm ini})}{d\cos\theta_W}\; [\PB]$}}
\put(262.8,441.6){\makebox(0,0)[rb]{\footnotesize one loop}}
\put(262.8,417.6){\makebox(0,0)[rb]{\footnotesize $\delta\kappa_Z = 0.1$}}
\put(262.8,393.6){\makebox(0,0)[rb]{\footnotesize $\lambda_Z = 0.1$}}
\put(262.8,369.6){\makebox(0,0)[rb]{\footnotesize $\xi_Z = 0.1$}}
\end{picture}
}
\caption[]{The effect on $d\sigma/d\cos\theta_W$ of one-loop corrections and
anomalous couplings at $\sqrt{s} = 174\GeV$.  Initial state radiation (in a
cone of $10^\circ$) has been taken into account unless otherwise stated.}
\label{fig:ano}
\end{figure}

The values used for the parameters were $m_Z = 91.176\GeV$,  $m_W =
80.152\GeV$, $m_t = 174\GeV$, $m_H=100\GeV$.  We assumed unpolarized beams of
$87\GeV$ each and used the $G_\mu$ renormalization scheme.  The results are
shown in Fig.\ \ref{fig:ano}.  One sees that the main effect of the radiative
corrections is described fairly well by a structure function in a cone of
$10^\circ$.  However, the difference is still of the same order of magnitude
as the deviations caused by a variation of $0.1$ in a single anomalous $WWZ$
coupling.

\section{Summary}

We have presented an event generator for $e^+e^-\to W^+W^-$ which includes
\begin{itemize}
\item full $\order(\alpha)$ corrections, including hard bremsstrahlung, both
in the $\alpha$ as in the $G_\mu$ scheme,
\item two-loop leading log effects for the initial state collinear
bremsstrahlung,
\item beam polarization (both transverse and longitudinal),
\item polarization of the $W$'s,
\item anomalous couplings of the triple gauge boson vertices.
\end{itemize}
It works in the energy range from threshold to 500 GeV (above this there are
numerical instabilities, especially in the forward region).  The generation
speed is of the order of 5 points per second on a workstation.  The main
limitation is the lack of offshell effects; we are working on this problem.




\appendix
\section{Installation}

The whole package has been written in Fortran 77 with some extensions and has
been tested on Sun\footnote{We had problems with the optimizer on Solaris},
DEC {\Large$\alpha$}, NeXT and HP\footnote{Do not use {\tt +O2}, change {\tt
xor} to {\tt ieor} in axo/ranf.f, line 498 and include {\tt flush.hp.o}}
workstations.  Due to the required range in floating point numbers,
$(m_e/m_W)^{20}$, it does not run as is on non-IEEE computers like VAX
\footnote{One could use {\tt GFLOAT}, but this normally causes problems with
libraries} and IBM mainframes.

The program consists of the following parts:
\begin{itemize}
\item two main programs; one reads its data from a file {\tt eeww.dat} and
calls {\tt vegas} or {\tt simplemc} to generate the events as a stand-alone
program; the other demonstrates the use of {\tt axmc}, which can more easily
be tied into standard libraries,
\item the switchyard routine {\tt wwmc}, corresponding to $f(\vec{x})$ in Eq.\
(\ref{eq:sum}), which calls {\tt virt} or {\tt hard},
\item the main routines with all the physics formul\ae,
\item the library {\tt axo.a}, which contains the {\tt vegas}
\cite{LepageVEGAS}, {\tt simplemc}, {\tt axmc} and supporting routines for
integration and event generation,
\item the tensor reduction library {\tt aa.a} and
\item the library of scalar functions {\tt ff.a} \cite{NewAlgorithms,FFguide}.
\end{itemize}
The whole package can be obtained with anonymous ftp from {\tt pss058.psi.ch}
in the subdirectory {\tt /pub/eeww.mc}, or as a {\tt gzip}'d {\tt tar} file.
The integration and event generation routines can easily be replaced by other
packages.


\section{Usage}

There are two ways to generate events.  The program {\tt eeww} reads its input
data from a file {\tt eeww.dat}, which it expects in the current directory.
This file is read a line at a time, with optional lines being skipped over.
An example file is shown below.

{\small
\begin{verbatim}
Standard    comment line
1           order to which the computation is done: 0 or 1
gmu         renormalization scheme: either 'alpha' or 'gmu'
91.176      mZ Z boson mass
80.152      mW W boson mass
130.0       mt top quark mass
100.0       mH Higgs boson mass
2   1d-3    IR: 0:normal cutoff, 1:Veltman, 2:computed, 3:rescale; cutoff/scale
2   10.     extra initial state coll. radiation: 0:not, 2:included; cone angle
polarized   electron polarization: unpolarized,lefthanded,righthanded,polarized
0. 0. .9 .9 90. 90. if polarized, polarization matrix (plm,plp,ptm,ptp,fim,fip)
polarized   W- polarization: unpolarized, polarized, transverse or longitudinal
polarized   W+ polarization: unpolarized, polarized, transverse or longitudinal
standard    anomalous couplings?  standard or nonstandard.
  .1 .1 .1  if nonstandard:      dkapg, lamg, xig
0 .1 .1 .1  if nonstandard: d1z, dkapz, lamz, xiz
1           number of CMS energies
200.        CMS energies
simple      method: one of 'vegas' or 'simple'
 1          simple: number of points points to generate
\end{verbatim}
}

The other method, useful for connecting with other programs, is demonstrated
by {\tt eewwax}.  This interface consists of an initialisation call, a call
which generates exactly one event, and an exit routine.  The parameters
corresponding to the data file are passed as arguments.

The events are output in the routine {\tt wweven}, which would be the point
where the $W$'s are decayed and the event analyzed.  A sample routine calling
{\tt jetset} \cite{jetset73} is included.  It has access to arrays of
four-momenta, polarization vectors and strings, and particle identification
codes.  Please refer to the comments in this files for details.

\section{Sample output}

The output of the program with the demo routine {\tt wwfeve}, given the input
file reproduced above, is shown below.  Please contact the authors if you have
problems reproducing this or other problems.

{\footnotesize
\begin{verbatim}
 =======================================================
 =                                                     =
 = EEEEE EEEEE W   W W   W  An order(alpha) generator  =
 = E     E     W   W W   W  for W pair production in   =
 = EEEE  EEEE  W   W W   W  electron positron events.  =
 = E     E     W W W W W W  J.Fleischer,F.Jegerlehner  =
 = EEEEE EEEEE  W W   W W   K.Kolodziej,GJ.v.Oldenborgh=
 =                                                     =
 =======================================================
 =            V E R S I O N    1 . 0                   =
 =======================================================
 ====================================================
     FF, a package to evaluate one-loop integrals
 written by G. J. van Oldenborgh, NIKHEF-H, Amsterdam
 ====================================================
 for the algorithms used see preprint NIKHEF-H 89/17,
 'New Algorithms for One-loop Integrals', by G.J. van
 Oldenborgh and J.A.M. Vermaseren, published in
 Zeitschrift fuer Physik C46(1990)425.
 ====================================================
 ffinit: precx =     4.4408920985006D-16
 ffinit: precc =     4.4408920985006D-16
 ffinit: xalogm =     4.9406564584125-324
 ffinit: xclogm =     4.9406564584125-324
Standard    comment line
 eeww: order(alpha) calculation
 eeww: working in the Gmu scheme
 eeww: using masses:
       mZ   =     91.176000000000
       mW   =     80.152000000000
       mtop =     130.00000000000
       mH   =    100.000000000000
 eeww: will compute the cutoff myself
 eeww: including the 2-loop leading log hard and
       exponentiated soft photon effects in cones
       of    10.0000000000000 degrees around the
       beam directions
 eeww: longitudinal polarization electron =   0.
       longitudinal polarization positron =   0.
       transverse   polarization electron =    0.90000000000000
       in direction     90.000000000000 degrees
       longitudinal polarization positron =    0.90000000000000
       in direction     90.000000000000 degrees
 eeww: generating all W- polarization states
 eeww: generating all W+ polarization states
 eeww: using standard W couplings
 eeww: running simple event generator
 eeww: CMS energy now is     200.00000000000 GeV
 ranf: using R250 generator
 maxweight increased to     22.968470711021 at event   -214
 NPOIN: warning: D4 is not yet supported
 NPOIN: warning: B1' seems also not yet supported
 ffxdbd: IR divergent B0', using cutoff     1.0000000000000D-24
 ffxdbd: using IR cutoff delta = lam^2 =     1.0000000000000D-24
 ffxc0i: infra-red divergent threepoint function, working with a cutoff
     1.0000000000000D-24
 ffzdbd: using IR cutoff delta = lam^2 =     1.0000000000000D-24
 maxweight increased to     37.140069173089 at event   -212
 maxweight increased to     37.296235986855 at event   -4
 simplemc: using as maximum weight:     44.755483184226
 #  id    E      px     py      pz      eps_0  eps_x  eps_y  eps_z   mass
polarization
 1  11  99.431   0.000  0.000  99.431   0.000  0.000  0.000  0.000   0.001
POLARIZED
 2 -11  99.431   0.000  0.000 -99.431   0.000  0.000  0.000  0.000   0.001
POLARIZED
 3 -24  99.117  19.937  2.770  54.723   0.000 -0.138  0.990  0.000  80.152 Y
TRANSVERSE
 4  24  99.748 -19.937 -2.770 -55.858   0.002  0.931  0.129 -0.342  80.152 X
TRANSVERSE
 6  22   1.135   0.000  0.000   1.135   0.000  0.000  0.000  0.000   0.000
 7  22   0.001   0.000  0.000  -0.001   0.000  0.000  0.000  0.000   0.000



                    The Lund Monte Carlo - JETSET version 7.3
                    **  Last date of change:  14 Jun 1991  **

 cross section using weights  =     18.732903091335
 cross section using 0 or 1   =     44.755483184226
 generated   1 points
 acceptance rate =    100.0000 %
 weight of each accepted event=     18.732903091335
 eeww: energy =     200.00000000000 GeV
       c.s. =     18.732903091335 +/-     18.732903091335
       chi2 =   0./DOF

total number of errors and warnings
===================================
fferr: no errors

\end{verbatim}
}

\end{document}